\documentclass[prc,twocolumn,showpacs]{revtex4}

\usepackage{graphicx}
\usepackage{dcolumn}
\usepackage{bm}

\begin{document}

\title{$\beta$-decay properties of Zr and Mo neutron-rich isotopes}

\author{P. Sarriguren$^{1}$}
\email{sarriguren@iem.cfmac.csic.es}
\author {J. Pereira$^{2}$}
\affiliation{
$^{1}$ Instituto de Estructura de la Materia, CSIC, Serrano
123, E-28006 Madrid, Spain \\
$^{2}$ National Superconducting Cyclotron Laboratory, MSU,
East Lansing, MI-48823, US}


\date{\today}

\begin{abstract}

Gamow-Teller strength distributions, $\beta$-decay half-lives, and
$\beta$-delayed neutron emission are investigated in neutron-rich
Zr and Mo isotopes within a deformed quasiparticle random-phase
approximation. The approach is based on a self-consistent Skyrme
Hartree-Fock mean field with pairing correlations and residual
separable particle-hole and particle-particle forces. Comparison
with recent measurements of half-lives  stresses the important role
that nuclear deformation plays in the description of  $\beta$-decay
properties in this mass region.

\end{abstract}


\maketitle

\section{Introduction}

The rapid neutron-capture process (r process) is considered as the
main nucleosynthesis mechanism responsible for the production of heavy
neutron-rich nuclei and for the existence of about half of the nuclei
heavier than iron \cite{bbhf,cowan91}. Although the astrophysical sites
for this process are still controversial, it takes place in scenarios
characterized by very high neutron densities. The path that
nucleosynthesis follows involves neutron-rich isotopes, which can be
far away from the valley of $\beta$-stability.
The most relevant nuclear properties to describe the r process are 
the nuclear masses and the $\beta$-decay properties 
\cite{cowan91,kra93} ---namely, the $\beta$-decay half-lives ($T_{1/2}$) 
and the  $\beta$-delayed neutron-emission probabilities ($P_n$). 
Nuclear masses define the possible r-process paths near the neutron 
drip-lines. The $T_{1/2}$ values of r-process waiting-point nuclei 
determine the pre freeze-out isotopic abundances and the speed of 
the process towards heavier elements, as well as the r-process time 
scale. The $P_n$ values of r-process isobaric nuclei define the 
decay path towards stability following the freeze-out, and provide a 
source of late-time neutrons.

A reliable nuclear physics description of the properties of the
extremely neutron-rich nuclei along the r-process path is needed to
interpret the astrophysical observations and to model and simulate
properly the r process. The quality of the nucleosynthesis modeling
is directly affected by the quality of the nuclear structure input.
Unfortunately, most of the nuclear properties of relevance for the
r process are experimentally unknown, although much effort is being
done recently, and therefore theoretical predictions must be 
considered. 
Such calculations are particularly challenging in the very exotic 
regions of interest, since they involve extrapolations using well 
established nuclear-structure models that have been properly tuned 
to account mostly for the properties of nuclei in the valley of 
stability.
In particular, the shell structure of neutron-rich drip-line nuclei
is still unknown to a large extent. Significant isospin dependence
of shell effects in medium-mass and heavy nuclei has been predicted
\cite{doba94,doba96,ots06}. 
It has been found that the shell gaps dramatically decrease near the 
neutron drip-lines because of continuum effects and a quenching of 
shell effects is apparent.

As a matter of fact, nuclear structure properties of nuclei far from 
stability where no experiments exist for direct comparison, can be 
tested by exploring their influence on the solar r-process abundance 
patterns. As an example, the agreement with the observed r-process 
abundances in the $A\sim 120$ mass region is manifestly improved
\cite{chen95,pfeiffer97,pfeiffer01,sun08} when using nuclear structure 
models that include a shell quenching effect at 
$N=82$ \cite{doba96,ETFSIQ}.

In this work we focus our attention on the mass region of neutron-rich
$A\sim 100-110$ nuclei, which is of great interest for the astrophysical
r process.  In addition, neutron-rich isotopes in this mass region are
known \cite{wood92} to be interesting examples where the equilibrium
shape of the nucleus is rapidly changing and shape coexistence is
present with competing prolate, oblate, and spherical shapes at close
energies (see e.g. \cite{cas85} for a general review).

In a recent publication \cite{pereira}, the $\beta$-decay properties
of some neutron-rich Zr and Mo isotopes were measured for the first
time. The data were interpreted in terms of the quasiparticle
random-phase approximation (QRPA) \cite{moller1,moller2,moller3,moller03},
using nuclear shapes and nuclear masses derived from the finite-range 
droplet model (FRDM) \cite{FRDM} and the latest version of the 
finite-range liquid-drop model (FRLDM) \cite{moller08}, which also 
includes triaxial deformation.
QRPA calculations for neutron-rich nuclei have also been performed
within different approaches, such as the Hartree-Fock-Bogoliubov
(HFB) \cite{engel}, continuum QRPA with either the extended 
Thomas-Fermi plus Strutinsky integral (ETFSI) method \cite{borzov1} 
or based on density functionals \cite{borzov2,borzov3}, and the
relativistic mean field (RMF) approach \cite{niksic2005}, just to
mention some of the recent publications, all of them for spherical
nuclei. However, the mass region of concern here requires nuclear
deformation as a relevant degree of freedom to characterize the
nuclear structure involved in the calculation of the $\beta$-strength
functions. The deformed QRPA formalism has been developed in Refs.
\cite{moller1,moller2,moller3,homma,hir1,hir2}, where phenomenological 
mean fields based on Nilsson or Woods-Saxon potentials were used as 
a starting basis. In this work we investigate the decay properties of 
neutron-rich even-even Zr and Mo isotopes within a deformed 
self-consistent Hartree-Fock (HF) mean field formalism with Skyrme 
interactions and pairing correlations in BCS approximation. 
Residual spin-isospin interactions are also included in the 
particle-hole and particle-particle channels and are
treated in QRPA \cite{sarri1,sarri2}.

The paper is organized as follows. In Sec. \ref{sec2} a brief review 
of the theoretical formalism is presented. Sec. \ref{results} contains 
the results obtained within this approach for the potential energy 
curves, Gamow-Teller (GT) strength distributions, and $\beta$-decay 
half-lives. Sec. IV summarizes the main conclusions.

\section{Theoretical Formalism}
\label{sec2}

In this section we show briefly the theoretical framework used in this
paper to describe the $\beta$-decay properties in Zr and Mo neutron-rich
isotopes. More details of the formalism can be found in 
Refs. \cite{sarri1,sarri2}.
The method consists of a self-consistent formalism based on a deformed
Hartree-Fock mean field obtained with Skyrme interactions, including
pairing correlations. The single-particle energies, wave functions, and
occupation probabilities are generated from this mean field.
In this work we have chosen the Skyrme force SLy4 \cite{sly4} as a
representative of the Skyrme forces. This particular force includes
some selected properties of unstable nuclei in the adjusting procedure
of the parameters. It is one of the more successful Skyrme forces
and has been extensively studied in the last years.

The solution of the HF equation is found by using the formalism 
developed in Ref. \cite{vautherin}, assuming time reversal and axial 
symmetry. The single-particle wave functions are expanded in terms 
of the eigenstates of an axially symmetric harmonic oscillator in 
cylindrical coordinates, using twelve major shells. The method also 
includes pairing between like nucleons in BCS approximation with 
fixed gap parameters for protons and neutrons, which are determined
phenomenologically from the odd-even mass differences through a
symmetric five term formula involving the experimental binding
energies \cite{audi} when available. 
In those cases where experimental information for masses is still 
not available, we have used the same pairing gaps as the closer 
isotopes measured. 
The pairing gaps for protons ($\Delta_p$) and neutrons ($\Delta_n$)
obtained in this way are roughly around 1 MeV. The corresponding pairing 
strengths $G_p$ and $G_n$ calculated from the gap equation depend
sensitively on the mass region, single-particle spectrum, and active
window for pairing. For typical values of the cutoffs of about 5 MeV
around the Fermi level, one obtains  $G_p\sim 0.25$ MeV and 
$G_n\sim 0.30$ MeV.
It is worth noticing that, although the BCS formalism leads to an 
unphysical neutron gas surrounding the nucleus near the drip line, 
the approximation is still valid in the region considered here, where 
the pairing gaps are still much lower than the Fermi energies.


The potential energy curves (PEC) are analyzed as a function of
the quadrupole deformation $\beta$, 

\begin{equation}
\beta = \sqrt{\frac{\pi}{5}}\frac{Q_0}{A\langle r^2 \rangle}\,
\label{beta_quadru}
\end{equation}
written in terms of the mass quadrupole moment $Q_0$ and the mean 
square radius $\langle r^2 \rangle$. For that purpose, constrained 
HF calculations are performed with a quadratic constraint 
\cite{constraint}. The HF energy is minimized under the constraint 
of keeping fixed the nuclear deformation. Calculations for GT 
strengths are performed subsequently for the equilibrium shapes of 
each nucleus, that is, for the solutions, in general deformed, for 
which minima are obtained in the energy curves. Since decays 
connecting different shapes are disfavored, similar shapes are 
assumed for the ground state of the parent nucleus and for all
populated states in the daughter nucleus.
The validity of this assumption was discussed for example in 
Refs. \cite{moller1,homma}. In our particular case, for SLy4 and 
neutron-rich Zr and Mo isotopes, the ground-state deformation of the
even-even parents (Zr,Mo) and of the corresponding $\beta$-decay 
odd-odd daughters (Nb,Tc) are practically the same, as it can 
be seen in Ref. \cite{web_sly4}.

To describe GT transitions, a spin-isospin residual interaction is
added to the mean field and treated in a deformed proton-neutron QRPA
\cite{sarri1,sarri2,moller1,moller2,moller3,homma,hir1,hir2}.
This interaction contains two parts, particle-hole ($ph$) and 
particle-particle ($pp$). The interaction in the $ph$ channel is 
responsible for the position and structure of the GT resonance 
\cite{sarri1,sarri2,homma} and it can be derived consistently from 
the same Skyrme interaction used to generate the mean field, through 
the second derivatives of the energy density functional with respect 
to the one-body densities. The $ph$ residual interaction is finally 
expressed in a separable form by averaging the Landau-Migdal resulting 
force over the nuclear volume, as explained in Refs. \cite{sarri1,sarri2}. 
The coupling strength is given by  $\chi ^{ph}_{GT}=0.15$ MeV.
The $pp$ part is a neutron-proton pairing force in the $J^\pi=1^+$ 
coupling channel, which is also introduced as a separable force 
\cite{hir1,hir2}.
The strength of the $pp$ residual interaction in this theoretical
approach is not derived self-consistently from the SLy4 force 
used to obtain the mean field basis, but nevertheless it has 
been fixed in accordance to it. This strength is usually fitted to 
reproduce globally the experimental half-lives. Various 
attempts have been done in the past to fix this strength, arriving 
to expressions such as $\kappa ^{pp}_{GT}=0.58/A^{0.7}$ in Ref.
\cite{homma}, which depend on the model used to describe the mean
field, Nilsson model in the above reference. Our work in the past 
(see Refs.  \cite{sarri09_1,sarri09_2} and references therein),
based on the Skyrme force SLy4 leads us to consider the value  
$\kappa ^{pp}_{GT} = 0.03$ MeV as a reasonable choice in this 
mass region, improving the agreement with the experimental
half-lives.

The proton-neutron QRPA phonon operator for GT excitations in
even-even nuclei is written as

\begin{equation}
\Gamma _{\omega _{K}}^{+}=\sum_{\pi\nu}\left[ X_{\pi\nu}^{\omega _{K}}
\alpha _{\nu}^{+}\alpha _{\bar{\pi}}^{+}+Y_{\pi\nu}^{\omega _{K}}
\alpha _{\bar{\nu}} \alpha _{\pi}\right]\, ,  \label{phon}
\end{equation}
where $\alpha ^{+}\left( \alpha \right) $ are quasiparticle creation
(annihilation) operators, $\omega _{K}$ are the QRPA excitation 
energies with respect to the ground state of the parent nucleus, 
and $X_{\pi\nu}^{\omega _{K}},Y_{\pi\nu}^{\omega _{K}}$ the 
forward and backward amplitudes, respectively. For even-even nuclei 
the allowed GT transition amplitudes in the intrinsic frame connecting 
the QRPA ground state
$\left| 0\right\rangle \ \ \left( \Gamma _{\omega _{K}} \left| 0
\right\rangle =0 \right)$ to one-phonon states $\left| \omega _K
\right\rangle \ \ \left( \Gamma ^+ _{\omega _{K}} \left| 0
\right\rangle = \left| \omega _K \right\rangle \right)$,
are given by

\begin{equation}
\left\langle \omega _K | \sigma _K t^{\pm} | 0 \right\rangle =
\mp M^{\omega _K}_\pm \, ,
\label{intrinsic}
\end{equation}
where
\begin{eqnarray}
M_{-}^{\omega _{K}}&=&\sum_{\pi\nu}\left( q_{\pi\nu}X_{\pi
\nu}^{\omega _{K}}+ \tilde{q}_{\pi\nu}Y_{\pi\nu}^{\omega _{K}}
\right) , \\
M_{+}^{\omega _{K}}&=&\sum_{\pi\nu}\left(
\tilde{q}_{\pi\nu} X_{\pi\nu}^{\omega _{K}}+
q_{\pi\nu}Y_{\pi\nu}^{\omega _{K}}\right) \, ,
\end{eqnarray}
with
\begin{equation}
\tilde{q}_{\pi\nu}=u_{\nu}v_{\pi}\Sigma _{K}^{\nu\pi },\ \ \
q_{\pi\nu}=v_{\nu}u_{\pi}\Sigma _{K}^{\nu\pi},
\label{qs}
\end{equation}
$v'$s are occupation amplitudes ($u^2=1-v^2$) and
$\Sigma _{K}^{\nu\pi}$ spin matrix elements connecting neutron
and proton states with spin operators
\begin{equation}
\Sigma _{K}^{\nu\pi}=\left\langle \nu\left| \sigma _{K}\right|
\pi\right\rangle \, .
\end{equation}

The GT strength $B_{\omega}(GT^\pm)$ in the laboratory system for a 
transition  $I_iK_i (0^+0) \rightarrow I_fK_f (1^+K)$ can be obtained 
in terms of the intrinsic amplitudes in Eq. (\ref{intrinsic}) as
\begin{eqnarray}
B_{\omega}(GT^\pm )& =& \sum_{\omega_{K}} \left[ \left\langle \omega_{K=0}
\left| \sigma_0t^\pm \right| 0 \right\rangle ^2 \delta (\omega_{K=0}-
\omega ) \right.  \nonumber  \\
&& \left. + 2 \left\langle \omega_{K=1} \left| \sigma_1t^\pm \right|
0 \right\rangle ^2 \delta (\omega_{K=1}-\omega ) \right] \, ,
\label{bgt}
\end{eqnarray}
in $[g_A^2/4\pi]$ units. To obtain this expression, the initial and
final states in the laboratory frame have been expressed in terms of
the intrinsic states using the Bohr-Mottelson factorization \cite{bm}.

The excitation energy $E_{ex}$ referred to the ground state of the
odd-odd daughter nucleus is obtained by
subtracting the lowest two-quasiparticle energy $E_0$ from the 
calculated $\omega$ energy in the QRPA calculation,
$E_{ex} = \omega_{QRPA} - E_0$, where $E_0=(E_n+E_p)_0$ is the
sum of the lowest quasiparticle energies for neutrons and protons. 
The GT strength B(GT) will be plotted later versus  $E_{ex}$ in Figs.
\ref{fig_bgt_zr}, \ref{fig_bgt_mo}, \ref{fig_bgt_zr_qbeta}, and
\ref{fig_bgt_mo_qbeta}.

The $\beta$-decay half-life is obtained by summing all the allowed
transition strengths to states in the daughter nucleus with
excitation energies lying below the corresponding $Q$-energy, and
weighted with the phase space factors $f(Z,Q_{\beta}-E_{ex})$,

\begin{equation}
T_{1/2}^{-1}=\frac{\left( g_{A}/g_{V}\right) _{\rm eff} ^{2}}{D}
\sum_{0 < E_{ex} < Q_\beta}f\left( Z,Q_{\beta}-E_{ex} \right) B(GT,E_{ex}) \, ,
 \label{t12}
\end{equation}
with $D=6200$~s and $(g_A/g_V)_{\rm eff}=0.77(g_A/g_V)_{\rm free}$,
where 0.77 is a standard quenching factor that takes into
account in an effective way all the correlations \cite{hama_eff}
which are not properly considered in the present approach.
The bare results can be recovered by scaling the results in this
paper for $B(GT)$ and $T_{1/2}$ with the square of this quenching
factor. The $Q_{\beta^-}$ energy is given by

\begin{eqnarray}
Q_{\beta^-}&=& M(A,Z)-M(A,Z+1)-m_e  \\
&=& BE(A,Z) -BE(A,Z+1)+m_n-m_p-m_e \, , \nonumber
\end{eqnarray}
written in terms of the nuclear masses $M(A,Z)$ or nuclear binding
energies $BE(A,Z)$ and the neutron ($m_n$), proton ($m_p$), and
electron ($m_e$) masses.

The Fermi integral $f(Z,Q_{\beta}-E_{ex})$ is computed numerically for each value 
of the energy including screening and finite size effects, as explained 
in Ref. \cite{gove},

\begin{equation}
f^{\beta^\pm} (Z, W_0) = \int^{W_0}_1 p W (W_0 - W)^2 \lambda^\pm(Z,W)
{\rm d}W\, ,
\end{equation}
with

\begin{equation}
\lambda^\pm(Z,W) = 2(1+\gamma) (2pR)^{-2(1-\gamma)} e^{\mp\pi y}
\frac{|\Gamma (\gamma+iy)|^2}{[\Gamma (2\gamma+1)]^2}\, ,
\end{equation}
where $\gamma=\sqrt{1-(\alpha Z)^2}$ ; $y=\alpha ZW/p$ ; $\alpha$ is
the fine structure constant and $R$ the nuclear radius. $W$ is the
total energy of the $\beta$ particle, $W_0$ is the total energy
available in $m_e c^2$ units, and $p=\sqrt{W^2 -1}$ is the momentum
in $m_e c$ units.

This function weights differently the strength $B(GT)$
depending on the excitation energy. As a general rule  $f(Z,Q_{\beta}-E_{ex})$
increases with the energy of the $\beta$-particle and therefore
the strength located at low excitation energies contribute more
importantly to the half-life.

The probability for $\beta$-delayed neutron emission is given by

\begin{equation}
P_n = \frac{ {\displaystyle \sum_{S_n < E_{ex} < Q_\beta}f\left( Z,Q_{\beta}-E_{ex}
\right) B(GT,E_{ex}) }}
{{\displaystyle \sum_{0 < E_{ex} < Q_\beta}f\left( Z,Q_{\beta}-E_{ex} \right)
B(GT,E_{ex})}}\, ,
\label{pn}
\end{equation}
where the sums extend to all the excitation energies in the daughter
nuclei in the indicated ranges. $S_n$ is the one-neutron separation
energy in the daughter nucleus. In this expression it is assumed that
all the decays to energies above $S_n$ in the daughter nuclei always
lead to delayed neutron emission and then, $\gamma$-decay from
neutron unbound levels is neglected.
According to Eq.~(\ref{pn}), $P_{n}$ is mostly sensitive to 
the strength located at energies around $S_n$, thus providing
a structure probe complementary to $T_{1/2}$.

\section{Results}
\label{results}

In this section we start by showing the results obtained for the
potential energy curves in the isotopes under study. Then, we calculate
the energy distribution of the GT strength corresponding to the local
minima in the potential energy curves. After showing the predictions
of various mass models to the $Q_\beta$ and $S_n$ values for the
more unstable isotopes, where no data on these quantities are available, 
we calculate the $\beta$-decay half-lives and
discuss their dependence on the deformation.

In previous works \cite{sarri1,sarri2,sarri_wp,sarri3,sarri4,sarri5}
we have studied the sensitivity of the GT strength distributions to 
the various ingredients contributing to the deformed QRPA-like 
calculations, namely to the nucleon-nucleon effective force, to 
pairing correlations, and to residual interactions. We found different 
sensitivities to them. In this work, all of these ingredients have been 
fixed to the most reasonable choices found previously and mentioned 
above, including the quenching factor. Here, we mainly discuss effects 
of deformation, keeping in mind that the method provides the 
self-consistent deformations as well.

\begin{figure}[ht]
\centering
\includegraphics[width=80mm]{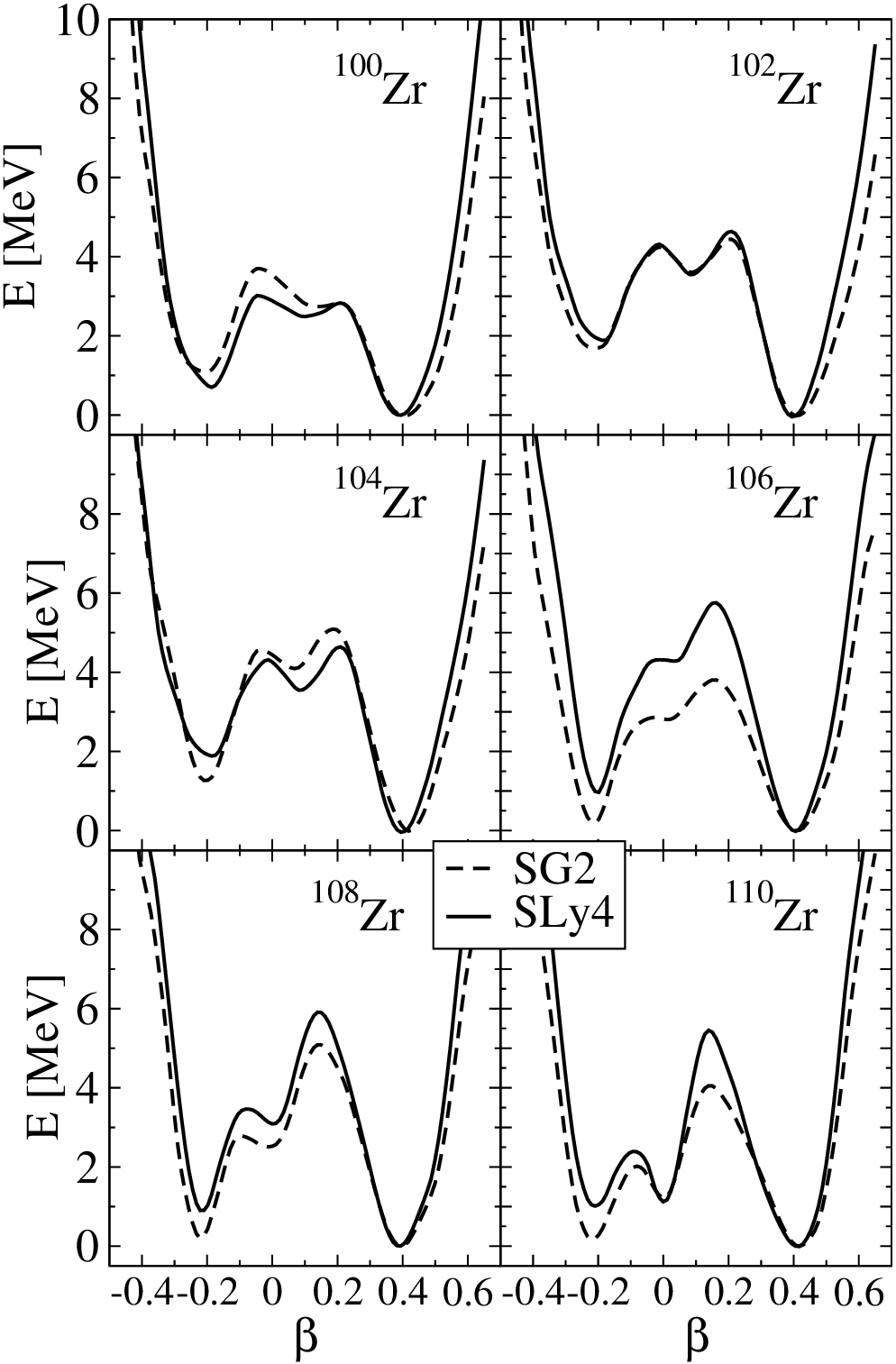}
\caption{Potential energy curves for even-even $^{100-110}$Zr isotopes
obtained from constrained HF+BCS calculations with the Skyrme forces
SG2 and SLy4.}
\label{fig_eq_zr}
\end{figure}

\begin{figure}[ht]
\centering
\includegraphics[width=80mm]{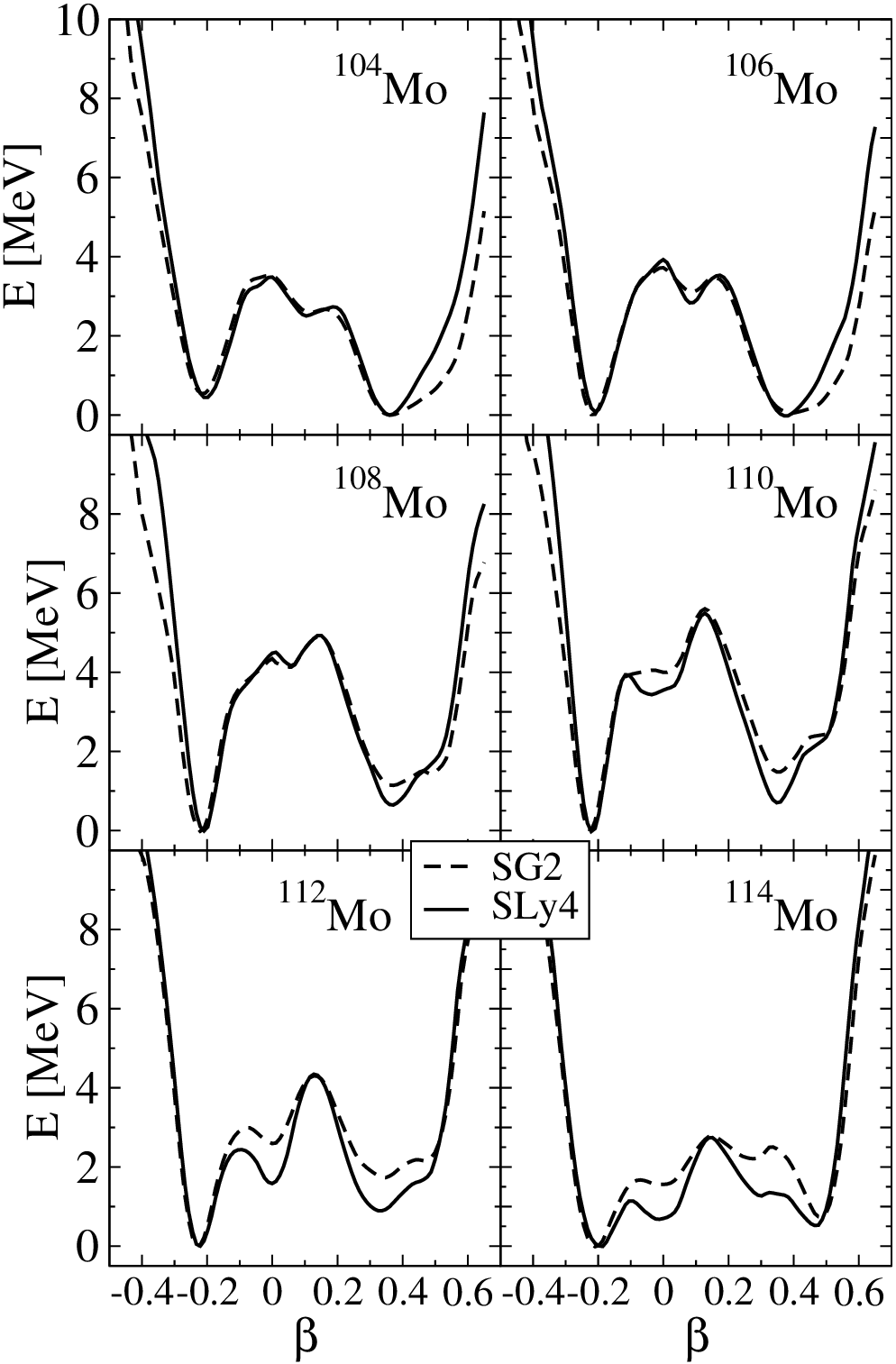}
\caption{Same as in Fig. \ref{fig_eq_zr}, but for  $^{104-114}$Mo 
isotopes.}
\label{fig_eq_mo}
\end{figure}

\subsection{Potential Energy Curves}

In Fig. \ref{fig_eq_zr}(\ref{fig_eq_mo}) we can see the potential energy
curves for the even-even $^{100-110}$Zr ($^{104-114}$Mo) isotopes. We show
the energies relative to that of the ground state plotted as a function
of the quadrupole deformation $\beta$. They are obtained from
constrained HF+BCS calculations with the Skyrme forces SG2 \cite{sg2}
and SLy4 \cite{sly4}. We observe that both forces produce very similar
results. In Fig.  \ref{fig_eq_zr} we see that the Zr isotopes exhibit
in all cases two well developed minima. The ground states are located
in the prolate sector at positive values of $\beta \approx 0.4$. We can 
also see oblate minima at higher energies located at $\beta \approx -0.2$. 
The two minima are separated by potential energy barriers varying from 
$E=3$ MeV, in the lightest $^{100}$Zr isotope, up to barriers of the 
order of 5 MeV in heavier isotopes. In the isotopes $^{108-110}$Zr a 
spherical local minimum is also developed.

Similar trends are observed in Fig.  \ref{fig_eq_mo} for the Mo isotopes.
We observe well developed oblate and prolate minima, which are
separated by barriers ranging from 3 MeV up to 5 MeV.  We get a prolate
ground state with an oblate minimum very close in energy in the
lightest isotope considered, $^{104}$Mo, a practically degenerate
oblate-prolate in $^{106}$Mo, and oblate ground states in heavier
isotopes with quadrupole deformations at $\beta \approx -0.2$
with prolate excited states at energies lower than 1 MeV. Again, the
heavier isotopes favor the appearance of a spherical configuration at
very low energies, resulting in an emergent triple
oblate-spherical-prolate shape coexistence scenario.

These results are in qualitative agreement with similar ones obtained
in this mass region from different  theoretical approaches. As an 
example of these methods we can mention the results obtained in
Ref. \cite{skalski97}, where this mass region was studied within a
macroscopic-microscopic approach based on an energy obtained from a
liquid drop or a FRLDM modified by a shell correction taken from a 
deformed Woods-Saxon potential. Zr isotopes from $N=60$ to $N=72$ were 
predicted to have well-deformed prolate ground states, while Mo isotopes 
suffered a shape transition from prolate shapes in the lighter 
neutron-rich isotopes ($N=62$) to oblate shapes in the heavier ones. 
Similarly, the deformations obtained in Ref. \cite{FRDM} from the FRDM 
and a folded-Yukawa single-particle microscopic model were in the range
$\beta= 0.36-0.38$ in the Zr isotopes considered in this work and
$\beta= 0.33-0.36$ in the Mo isotopes, except in the heavier 
$^{114}$Mo, where an oblate shape with $\beta\approx -0.25$ becomes the 
ground state. RMF calculations \cite{lala1,lala2} show ground-state
deformations in the range of $\beta= 0.36-0.40$ in the Zr isotopes,
while for Mo isotopes oblate ground states are obtained with parameters
of deformation between $\beta=-0.28$ and $\beta=-0.23$, except in the
lighter isotope $^{104}$Mo, where a prolate ground state  $\beta=0.336$
is found.  Calculations including rotational states in terms of the total 
Routhian surface (TRS), using non-axial Wood-Saxon potentials \cite{xu02}, 
predicted two coexisting prolate and oblate minima ($\beta \approx 0.35$ 
and $\beta \approx -0.2$) for $^{106-116}$Zr isotopes, where the prolate 
ground state becomes oblate beyond $^{110}$Zr. The same calculations 
showed oblate ($\beta \approx -0.22$) ground-states for $N > 68$ Mo 
isotopes.
Finally, similar results in the sense of competing oblate and prolate
shapes and emergence of spherical configurations in the heavier isotopes
are also obtained within the Hartree-Fock-Bogoliubov framework with the
finite-range effective Gogny interaction D1S \cite{hilaire}. Equilibrium
oblate ($\beta \approx -0.2$) and prolate ($\beta \approx 0.4$) coexistent
deformations were found in Ref. \cite{hilaire} practically at the same
energy in the Zr isotopes. In Mo isotopes an oblate shape
($\beta \approx -0.2$) is favored energetically with close prolate
($\beta \approx 0.4$) solutions. In both cases, Zr and Mo isotopes, a
spherical solution lows in energy and becomes almost degenerate with
the deformed solutions for the heavier isotopes $^{110}$Zr and 
$^{114}$Mo. 

Thus, a consistent theoretical picture emerges, which is  supported by
the still scarce experimental information available. Experimentally,
two coexisting deformed bands weakly admixed were found in $^{100}$Zr
\cite{mach1,mach2} from an analysis of  $B(E2)$ and $\rho (E0)$ and a 
two-level mixing model analysis. One of these bands is a highly deformed 
prolate yrast band ($\beta=0.34$), while the other is moderately deformed
($|\beta|=0.16$) and weakly mixed to the yrast by about $10\%$. The 
highly deformed band in $^{100}$Zr is nearly identical to the yrast band 
in $^{102}$Zr. 
Hill et al. \cite{hil91} have also discussed the possibility that the
$0_{2}^{+}$ level measured for $^{102}$Zr at 895~keV could be the head 
of a band with $|\beta|\approx 0.2$, similar to the S band of $^{100}$Zr.

Quadrupole moments were also determined \cite{urban} for rotational 
bands  in $^{98-104}$Zr isotopes and deformation parameters were deduced 
increasing gradually from $\beta=0.1$ at $N=56$  up to  $\beta=0.4$ at 
$N=64$. More recently \cite{goodin}, large deformations ($\beta=0.47(7)$) 
were extracted in  $^{104}$Zr  and in   $^{106}$Mo ($\beta=0.36(7)$) 
from the half-lives of their $2^+$ states. 
Spectroscopic studies of high-spin states of $^{100-104}$Zr and 
$^{102-108}$Mo have also been performed by Hua et al. \cite{hua04} 
within the particle-rotor model. According to these authors, the 
difference in signature splitting observed for the 5/2$^{-}$[532] 
band between the odd Zr and Mo isotopes could be attributed to the 
appearance of triaxiality in Mo isotopes.
As mentioned above, the formalism employed in the present study does 
not include non-axial deformation. Such limitation, however, has no
significant impact in the results discussed here. As an example, the 
inclusion of triaxiality in the last version of the FRLDM~\cite{moller08}
resulted in a small reduction of the $^{106,108}$Mo ground-state energies 
(of about 250~keV) at $\gamma$=17.5$^{\circ}$, with respect
to pure prolate shapes. Similarly, Xu et al. \cite{xu02} predict a 
$\gamma$-soft triaxial minimum for $^{108}$Mo.

\begin{figure}[ht]
\centering
\includegraphics[width=80mm]{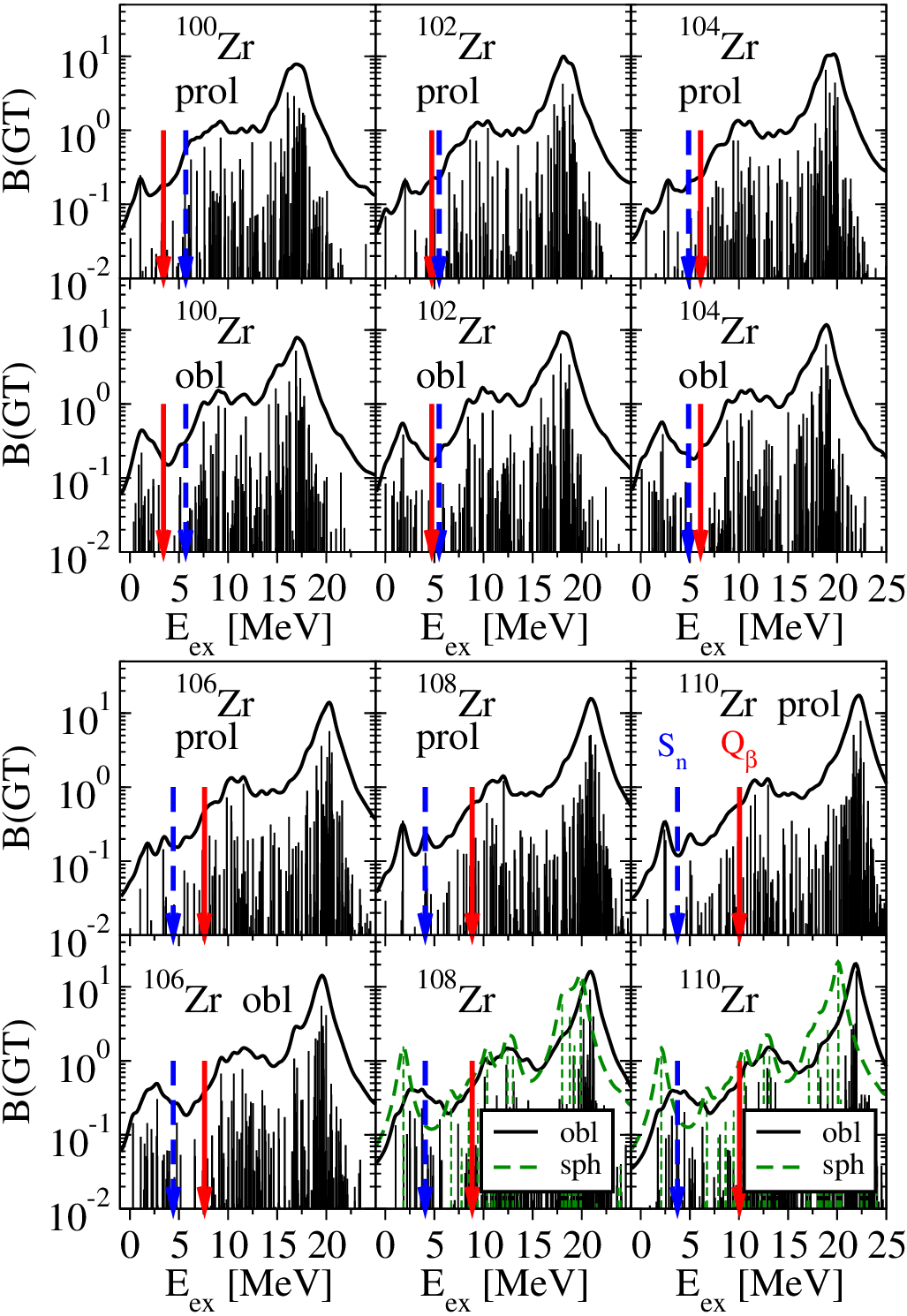}
\caption{(Color online) QRPA-SLy4 Gamow-Teller strength distributions 
for Zr isotopes as a function of the excitation energy in the daughter
nucleus. The calculations correspond to the various equilibrium
configurations found in the PECs curves.
$Q_\beta$ and $S_n$ values are shown by solid and dashed vertical arrows,
respectively. }
\label{fig_bgt_zr}
\end{figure}

\begin{figure}[ht]
\centering
\includegraphics[width=80mm]{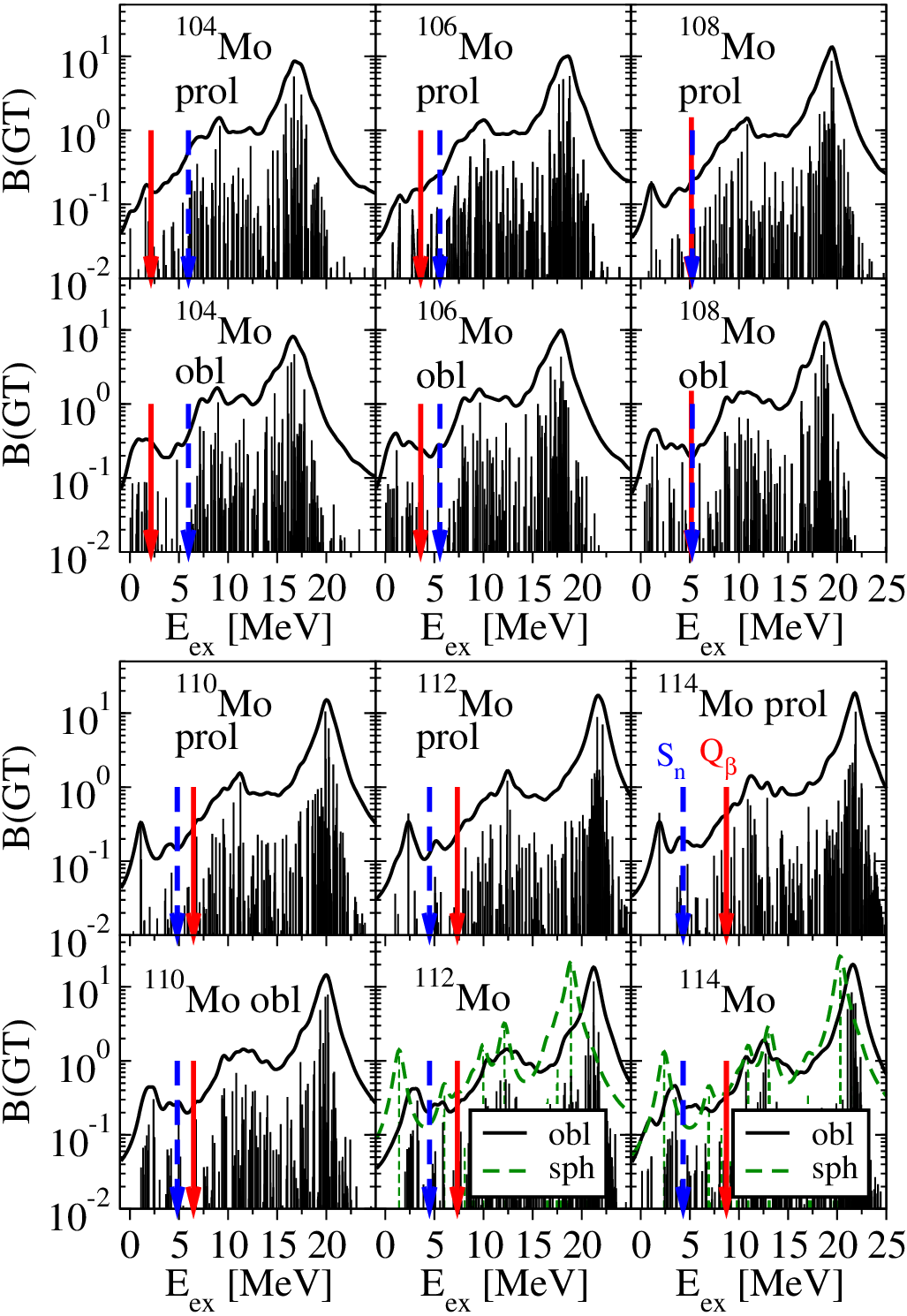}
\caption{(Color online) Same as in Fig. \ref{fig_bgt_zr}, but for Mo isotopes.}
\label{fig_bgt_mo}
\end{figure}

\subsection{Gamow-Teller strength distributions}

In the next figures, we show the results obtained for the energy
distributions of the GT strength corresponding to the
oblate-prolate-spherical equilibrium shapes for which we obtained
minima in the potential energy curves in Figs. \ref{fig_eq_zr} and
\ref{fig_eq_mo}. The results are obtained with the force SLy4, using 
constant pairing gaps extracted from the experimental masses (or 
systematics) and with residual interactions with the parameters written 
in Sec. \ref{sec2}.  The GT strength in $(g_A^2/4\pi)$ units, is 
plotted versus the excitation energy of the daughter nucleus and a 
quenching factor 0.77 has been included.

\begin{figure}[ht]
\centering
\includegraphics[width=80mm]{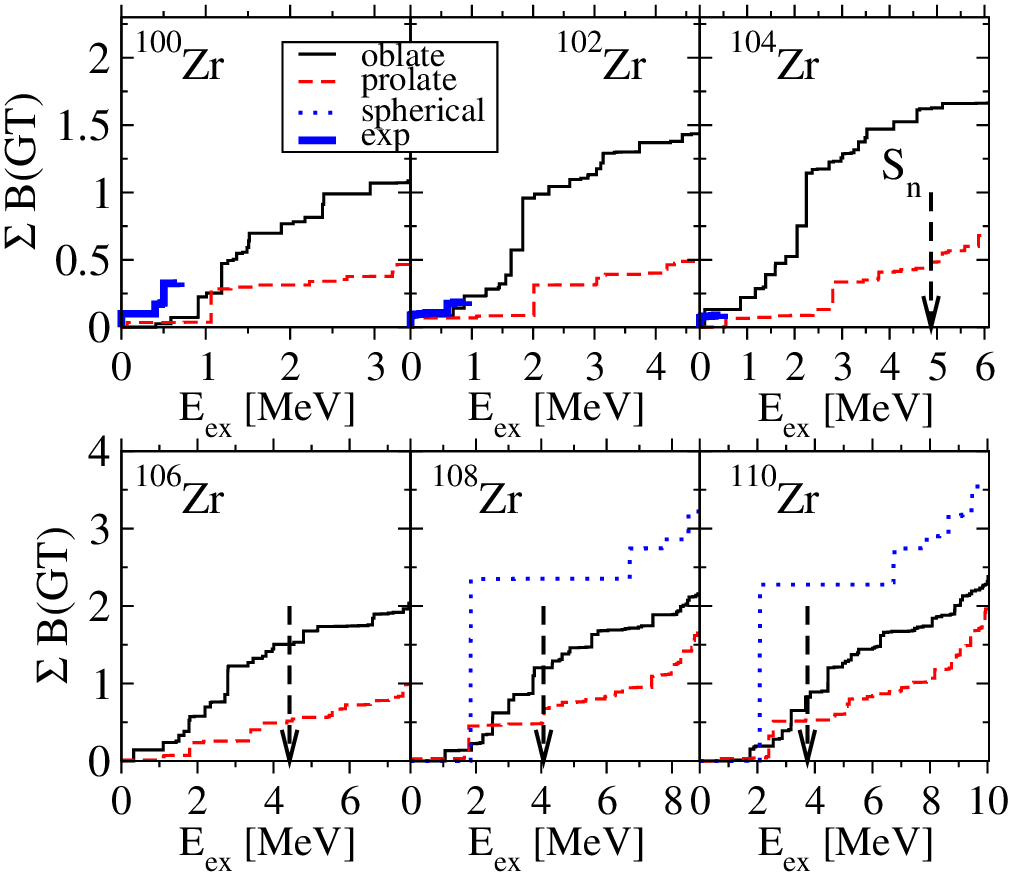}
\caption{(Color online) QRPA-SLy4 accumulated GT strengths in Zr isotopes 
calculated for 
the various equilibrium shapes. In each isotope the energy range considered
corresponds to its $Q_\beta$ value. $S_n$ values are shown by dashed vertical 
arrows.}
\label{fig_bgt_zr_qbeta}
\end{figure}

\begin{figure}[ht]
\centering
\includegraphics[width=80mm]{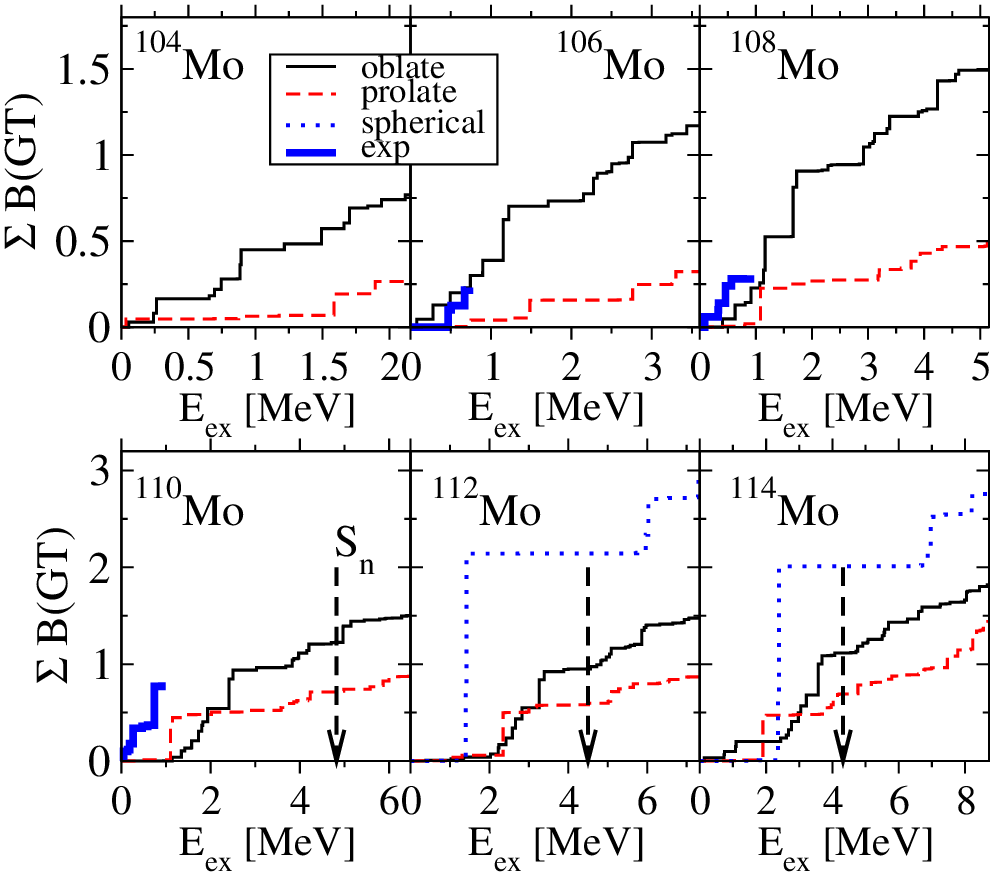}
\caption{(Color online) Same as in Fig. \ref{fig_bgt_zr_qbeta}, but for Mo 
isotopes.}
\label{fig_bgt_mo_qbeta}
\end{figure}

Figs. \ref{fig_bgt_zr} and \ref{fig_bgt_mo} contain the results for Zr
and Mo isotopes, respectively. We show the energy distributions of the
individual GT strengths together with continuous distributions obtained
by folding the strength with 1 MeV width Breit-Wigner functions. The
vertical arrows show the $Q_\beta$ and $S_n$ energies, taken from experiment
\cite{audi} or from the mass formula in Ref. \cite{DufloZuker} when data
are not available, as we shall explain later on.

The main characteristic of these distributions is the existence of a GT
resonance located at increasing excitation energy as the number of neutrons
$N$ increases. The total GT strength also increases with $N$, as it is
expected to fulfill the Ikeda sum rule. It is worth noticing that both
oblate and prolate shapes produce quite similar GT strength distributions
on a global scale. Even the spherical profiles are quite close to the
deformed ones. Nevertheless, the small differences among the various shapes 
at the low energy tails (below the $Q_\beta$) of the GT strength distributions
that can be appreciated because of the logarithmic scale, 
lead to sizable effects in the $\beta$-decay half-lives. In the next figures,
Fig. \ref{fig_bgt_zr_qbeta} for Zr isotopes and Fig. \ref{fig_bgt_mo_qbeta}
for Mo isotopes, we can see the accumulated GT strength plotted up to
the corresponding $Q_\beta$ energy of each isotope, which is the relevant
energy range for the calculation of the half-lives. Also shown by vertical
dashed lines are the $S_n$ energies when they are lower than  $Q_\beta$.
In this magnified scale one can appreciate the sensitivity of these
distributions to deformation and how measurements of the GT strength
distribution from $\beta$-decay can be a tool to get information about
this deformation, as it was carried out in Refs. \cite{exp_poirier,exp_nacher}.
The accumulated strength from the oblate shapes is in general larger than 
the corresponding prolate profiles. The spherical distributions have distinct 
characteristics showing always as a strong peak at an excitation energy 
around 2 MeV. The profiles from different shapes could be easily
distinguished experimentally from each other. This is specially true
in the case of the lighter isotopes $^{100-104}$Zr and  $^{104-108}$Mo,
where the differences are enhanced. These isotopes are in principle
easier to measure since they are the less exotic.

Experimental information on GT strength distributions in these isotopes
is only available in the energy range below 1 MeV for the isotopes
$^{106,108}$Mo \cite{jokinen}, $^{110}$Mo \cite{wang04}, and
$^{100,102,104}$Zr \cite{rinta07}. These data can be seen in Figs.
\ref{fig_bgt_zr_qbeta} and \ref{fig_bgt_mo_qbeta}, together with the
QRPA calculations.
Unfortunately, the energy region is still very narrow and represents
only a small fraction of the GT strength relevant for the half-life
determination. Clearly, more experimental information is needed
to get insight into the nuclear structure of these isotopes. 

\subsection{Half-lives and $\beta$-delayed neutron-emission probabilities}

The calculation of the half-lives in Eq. (\ref{t12}) involves the knowledge
of the GT strength distribution and of the $Q_\beta$ values. The calculation 
of the probability for $\beta$-delayed neutron emission $P_n$ in Eq. (\ref{pn})
involves also the knowledge of the $S_n$ energies. We use experimental values 
for $Q_\beta$ and $S_n$, which are taken from Ref. \cite{audi} or from the 
Jyv\"askyl\"a mass database \cite{jyvaskyla},  when available.
But in those cases where experimental masses are not available, one has to
rely on theoretical predictions for them.
There are a large number of mass formulas in the market obtained from
different approaches.

The strategy used in this work is first of all to compare with experiment
the predictions of some representative mass formulas in the mass region
where data are available.
According to their success in reproducing the $Q_\beta$ and $S_n$ energies,
we finally adopt the most convenient mass formula for extrapolations to
the unknown regions.

\begin{figure}[ht]
\centering
\includegraphics[width=80mm]{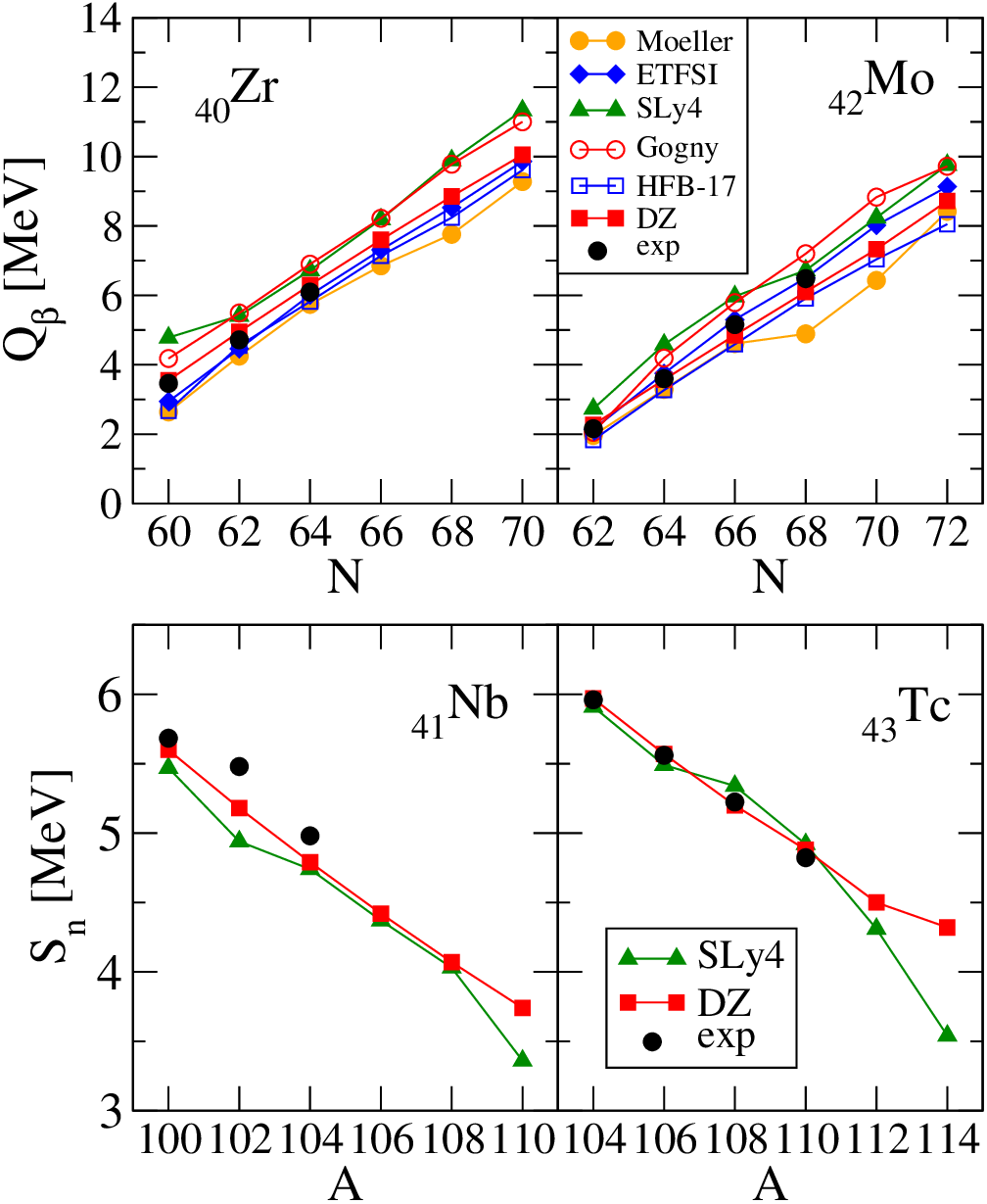}
\caption{(Color online) Experimental $Q_\beta$ and $S_n$ energies compared to the
predictions of various mass models.}
\label{fig_qbeta_sn}
\end{figure}

In Fig. \ref{fig_qbeta_sn} we show this comparison for six frequently 
used mass formulas. We use the model of P. M\"oller et al. \cite{FRDM} 
that belongs to a microscopic-macroscopic type of calculation. It 
contains a FRDM corrected by microscopic effects obtained from a deformed
single-particle model based on a folded Yukawa potential including 
pairing in the Lipkin-Nogami approach. Then we use the ETFSI model 
\cite{ETFSI}, which adopts a semi-classical approximation to the
Hartree-Fock method including full Strutinsky shell corrections and BCS
pairing correlations. The label SLy4 stands for the masses calculated
from the Skyrme force SLy4 with a zero-range pairing force and
Lipkin-Nogami obtained from the code HFBTHO \cite{mass_sly4} and
compiled in \cite{web_sly4}. The results under the label Gogny have
been obtained from the HFB calculations with the finite-range Gogny-D1S 
force \cite{hilaire}. The HFB-17 model is one of the most recent versions 
of the Skyrme HFB mass formulas introduced by the Brussels-Montreal group 
\cite{HFB17_1,HFB17_2}. As in the case of the previous cases, SLy4 and 
Gogny, this is a fully microscopic approach since it is based on an 
effective two-body nucleon-nucleon interaction.
The Duflo and Zuker (DZ) mass model \cite{DufloZuker} is written as
an effective Hamiltonian which contains two parts, a monopole term and
a multipole term. The monopole calculations are purely HF-type based on
single-particle properties while the multipole term acts as a residual
interaction and the calculation goes beyond HF. Its predictive power
has been recently checked \cite{mendoza} with a number of tests
probing its ability to extrapolate with very good results.
In this work we use the 10-parameter version of the mass formula
\cite{mendoza_pre}, which is a simplification of the more sophisticated
28-parameter mass formula in Ref. \cite{DufloZuker}.

In the upper panels of Fig. \ref{fig_qbeta_sn} we can see the experimental
$Q_\beta$ values (black dots) \cite{audi,jyvaskyla}, available for the
isotopes $^{100,102,104}$Zr and $^{104,106,108,110}$Mo. They are compared with
the predictions of the various mass models discussed above.  In the lower
panels we have the neutron separation energies $S_n$ corresponding to the
daughter isotopes of Nb and Tc, where we compare the measured energies 
(black dots) with the predictions of the DZ formula and SLy4 force.
We have selected for consistency the SLy4 predictions, but also 
the DZ mass formula as one of the the most suited formula
in this particular mass region. They agree pretty well with the measured 
values for both $Q_\beta$ and $S_n$ values. In what follows the results for 
half-lives and $P_n$ for the $^{106,108,110}$Zr and $^{112,114}$Mo will be 
obtained by using $Q_\beta$ and $S_n$ from SLy4 and DZ mass formula.

\begin{figure}[ht]
\centering
\includegraphics[width=80mm]{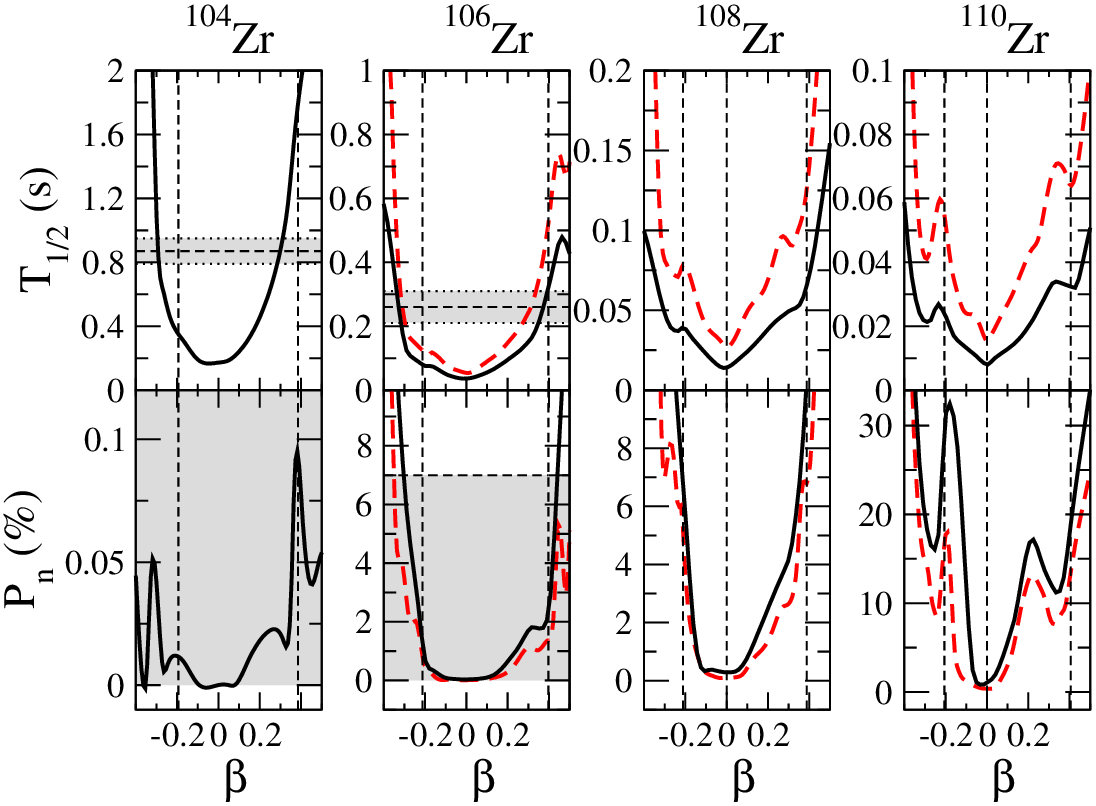}
\caption{(Color online) QRPA-SLy4 
$\beta$-decay half-lives and $P_n$ values for
Zr isotopes as a function of the quadrupole deformation $\beta$ compared
to experiment (shaded area). See text for more details.}
\label{fig_t_beta_zr}
\end{figure}

\begin{figure}[ht]
\centering
\includegraphics[width=80mm]{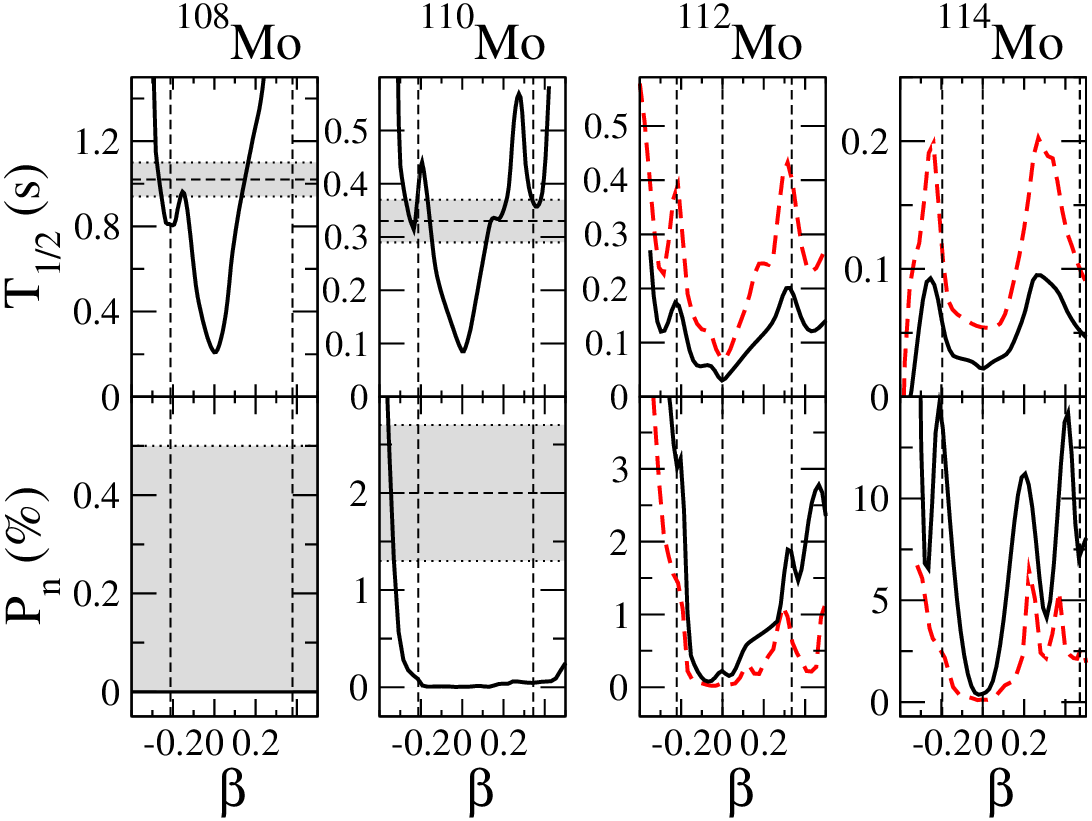}
\caption{(Color online) Same as in Fig. \ref{fig_t_beta_zr}, but for Mo isotopes.}
\label{fig_t_beta_mo}
\end{figure}

In Figs. \ref{fig_t_beta_zr} and \ref{fig_t_beta_mo} we can see the 
dependence of the half-lives $T_{1/2}$ and $P_n$ values with the 
quadrupole deformation $\beta$. 
Solid lines in the lighter isotopes ($^{104}$Zr and $^{108,110}$Mo)
correspond to QRPA-SLy4 calculations using experimental $Q_{\beta}$ 
and $S_n$. In the heavier isotopes ($^{106-110}$Zr and $^{112,114}$Mo), 
where there are no data for $Q_{\beta}$ and $S_n$, solid 
(dashed) lines correspond to QRPA-SLy4 calculations using $Q_{\beta}$ 
and $S_n$ from SLy4 (DZ). Experimental data are shown by 
horizontal dashed lines, where the shaded region in between corresponds 
to a 1-$\sigma$ confidence level. The vertical dashed lines show the
self-consistent quadrupole deformations for which we obtained the
equilibrium shape configurations (see Figs. \ref{fig_eq_zr} and
\ref{fig_eq_mo}).
The first evidence to mention is that a spherical approach to these
nuclei is far from the measured data, demanding a deformed treatment.

In Fig.  \ref{fig_t_beta_zr} we show the results for the isotopes
$^{104-110}$Zr. In the cases of $^{104,106}$Zr we reproduce the experimental
half-lives with oblate and prolate deformations very close to the
self-consistent ones. In the oblate case the calculation gives half-lives
lower than experiment, while the self-consistent prolate deformation
produces somewhat larger ones. Thus, the experiment would be reproduced either
by nuclear deformations which do not produce shapes at equilibrium
($\beta\approx \pm 0.3$) or by a mixing of the equilibrium deformations.
Interestingly, similar results were obtained in Ref. \cite{pereira} 
for $^{104}$Zr, from the analysis of the measured $\beta$-decay 
properties of $^{104}$Y in terms of quadrupole deformation $\epsilon_{2}$ 
of the daughter $^{104}$Zr (see \cite{FRDM} for a formal definition 
of the parameters $\epsilon_{2}$ and $\beta$). 
In that case, the quadrupole deformation needed to reproduce the measured 
half-life and $P_{n}$ value was $|\epsilon_{2}| \approx 0.25$, although 
oblate deformations were ruled out at that time. An important difference 
between the results shown in Fig. \ref{fig_t_beta_zr} and those discussed 
in Ref. \cite{pereira} is the abrupt increase seen in the latter for the 
$T_{1/2}$ and $P_{n}$ value for a near-spherical $^{104}$Zr. These large 
values were mostly produced by the location of the GT-populated 
$\pi$g$_{9/2}$$\otimes$$\nu$g$_{7/2}$ level at rather high energies in the 
spherical daughter $^{104}$Zr. This discrepancy emphasizes the sensitivity 
of $T_{1/2}$ and $P_{n}$ to the structure details of the mother/daughter 
nuclei. The experimental $P_n$ values are only upper limits, although 
they are much larger than the typical values obtained theoretically.
In the heavier isotopes there are no data and these results are then
useful to see the sensitivity to deformation of the predictions. The 
spherical minima in the heavier isotopes predict half-lives and $P_n$ 
values much lower than the corresponding values for deformed shapes.
In Fig.  \ref{fig_t_beta_mo} we have the results for the isotopes
$^{108-114}$Mo. In the case of $^{108}$Mo the half-life is reproduced
with the self-consistent oblate deformation, while the prolate one
generates too high half-lives. In the case of $^{110}$Mo the measured
half-life is well reproduced with both oblate and prolate equilibrium
deformations.  In the case of $^{108}$Mo the $P_n$ value is zero since
experimentally \cite{jyvaskyla} $S_n > Q_\beta$. For  $^{110}$Mo the
$P_n$ value is not reached by the calculations. As in the case of the
heavier Zr isotopes, the heavier Mo isotopes show that the half-lives
for the spherical minima are much smaller than the corresponding
half-lives for the self-consistent oblate and prolate shapes.
In general we observe that the half-lives ($P_n$ values) in the 
heavier Zr and Mo isotopes calculated with $Q_{\beta}$ and $S_n$ from 
SLy4 (solid lines) are lower (larger) than the results calculated with 
$Q_{\beta}$ and $S_n$ from DZ (dashed lines).

\begin{figure}[ht]
\centering
\includegraphics[width=80mm]{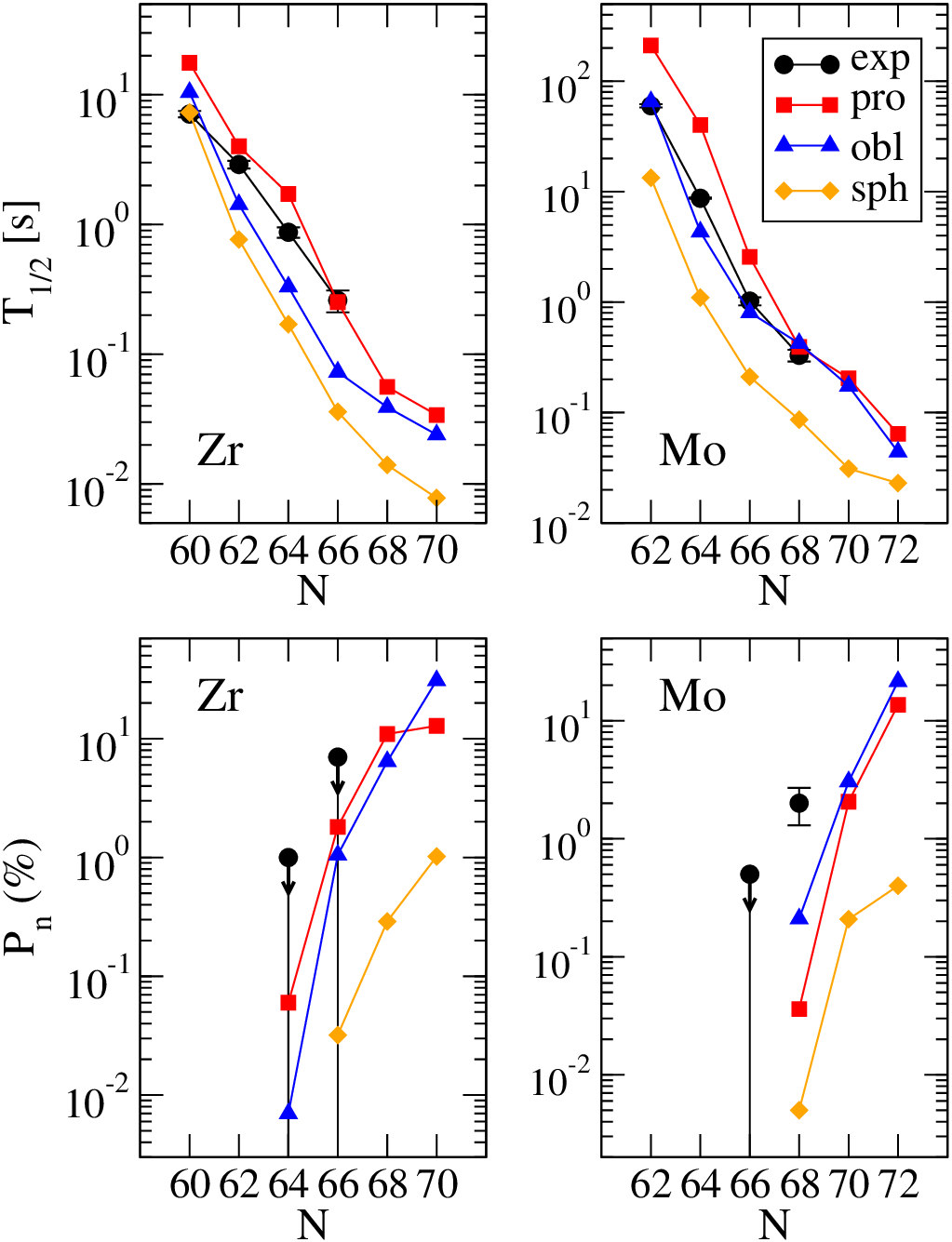}
\caption{(Color online) Measured  $\beta$-decay half-lives and $P_n$ values
for Zr and Mo isotopes compared to theoretical QRPA results
calculated from different shape configurations, using SLy4 to compute
$Q_\beta$ and $S_n$ in the heavier isotopes.}
\label{fig_t_pn_sly4}
\end{figure}

\begin{figure}[ht]
\centering
\includegraphics[width=80mm]{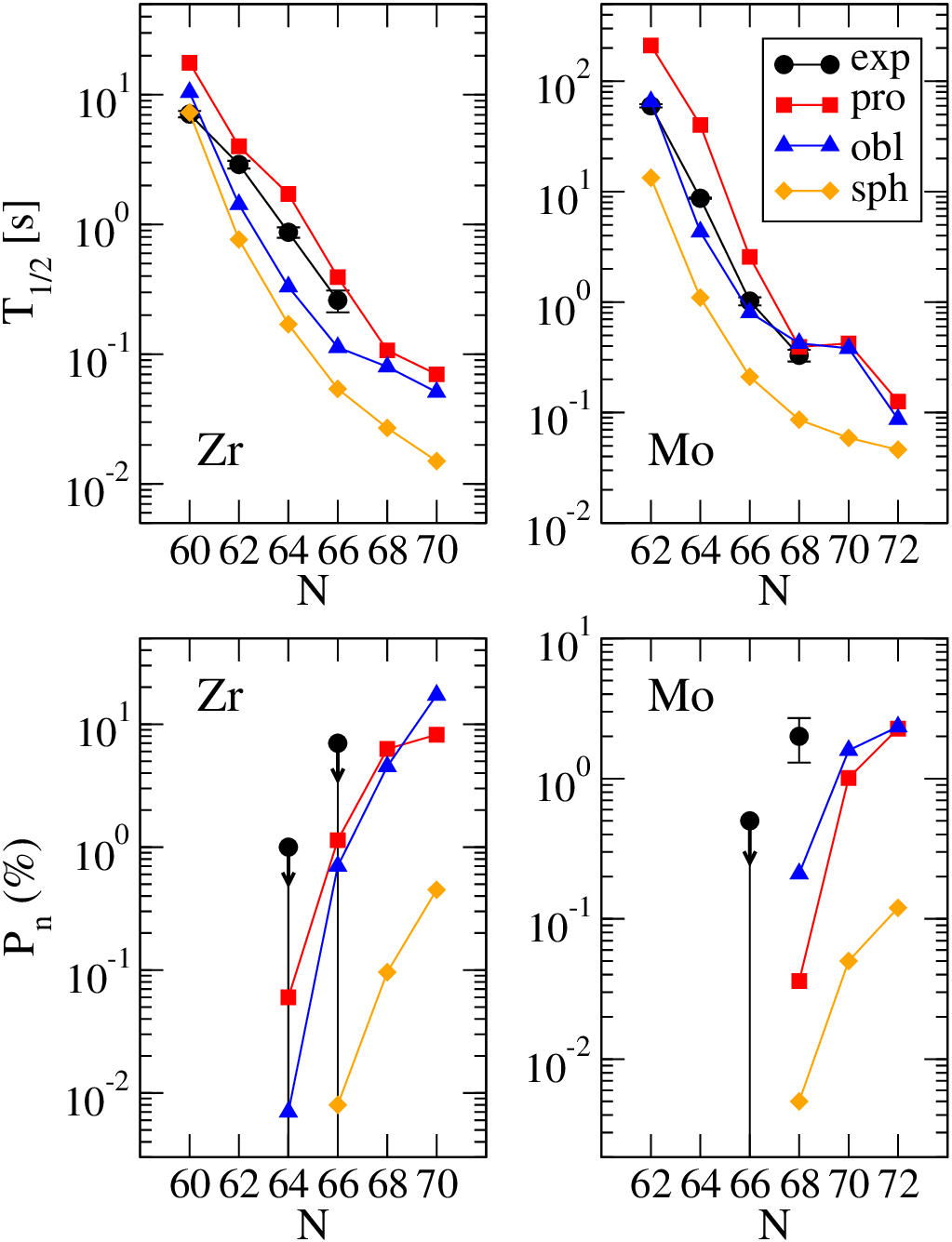}
\caption{(Color online) Same as in Fig. \ref{fig_t_pn_sly4}, but using DZ
instead of SLy4 to compute $Q_\beta$ and $S_n$ in the heavier isotopes.}
\label{fig_t_pn_dz}
\end{figure}

In Figs. \ref{fig_t_pn_sly4} and \ref{fig_t_pn_dz} we compare the measured 
$\beta$-decay half-lives
(upper panels) and $P_n$ values (lower panels) with the theoretical
results obtained with the oblate, prolate, and spherical equilibrium 
shapes. In the  $^{100-104}$Zr isotopes we use experimental $Q_\beta$ and 
$S_n$ values, while in the heavier  $^{106-110}$Zr isotopes we use SLy4
in Fig. \ref{fig_t_pn_sly4} and the DZ mass formula in Fig. \ref{fig_t_pn_dz}.
Similarly, for the  $^{104-110}$Mo isotopes we use experimental $Q_\beta$
and $S_n$ values, while for  $^{112-114}$Mo we use SLy4
in Fig. \ref{fig_t_pn_sly4} and the DZ mass formula in Fig. \ref{fig_t_pn_dz}.
In the case of Zr isotopes, we can see that the experimental half-life
is close to the oblate result in $^{100}$Zr and appear systematically
between the prolate and oblate calculations in the isotopes 
$^{102,104,106}$Zr.
One wonders whether such result could be explained by the coexistence 
of a highly-deformed prolate ground-state configuration with a moderately 
deformed minimum similar to that found in $^{100}$Zr \cite{mach1,mach2} and 
(more speculatively) in $^{102}$Zr \cite{hil91}. The results 
seem to indicate that such weakly-deformed intruder 
configurations may have an oblate character. In the heavier isotopes 
$^{108,110}$Zr the predictions of both oblate and prolate are very close 
to each other and much larger than the result obtained from spherical 
shapes. Measuring these half-lives and $P_{n}$ values will be a good
opportunity to check the role of spherical configurations in these exotic
nuclei, since the spherical components will lower the half-lives and 
$P_{n}$ values by factors about 5 and 15-50, respectively.

In the case of Mo isotopes, the experimental half-lives in 
$^{104,106,108,110}$Mo tend to favor the oblate theoretical results (which 
are indeed the ground states) over the prolate ones. In the heavier  
$^{112,114}$Mo isotopes, as in the case of the heavier Zr isotopes, 
oblate and prolate results are very similar and much larger than the 
spherical predictions, offering again a sensitive test to analyze the 
deformation of these heavy nuclei, for which spectroscopic measurements 
are more difficult.
Experimental $P_n$ values are only upper limits except for the case 
$^{110}$Mo, which is much larger than the calculations. This implies 
that the relative GT strength contained in the energy region below 
$S_n$ is overestimated theoretically and therefore the relative 
contribution coming from the strength above $S_n$ is too small. This 
can be seen in Fig. \ref{fig_bgt_mo_qbeta} for $^{110}$Mo, where the 
accumulated strength is practically flat above $S_n$.
The half-lives and $P_{n}$ values of the $A\sim 110$ nuclei, predicted 
here for spherical configurations, would have clear consequences in the 
calculation of r-process abundances. In particular, the abrupt reduction 
of the $P_{n}$ values may contribute to fill the artificial trough around 
$A=110$ predicted by current r-process nucleosynthesis models.
Furthermore, the confirmation of spherical shapes in these nuclei may be 
an indirect signature of the $N=82$ shell quenching since both phenomena 
are predicted by the SLy4 force used in our calculations.

\section{CONCLUSIONS}

In this paper we have studied the $\beta$-decay properties of neutron-deficient
Zr and Mo isotopes within a deformed QRPA approach based on mean
fields generated from self-consistent Skyrme Hartree-Fock calculations.
In particular, we have analyzed the experimental information on the
half-lives and $\beta$-delayed neutron-emission probabilities in the
neutron-rich $^{100-110}$Zr and $^{104-114}$Mo isotopes in terms of the
nuclear deformation.

We have shown that the measured half-lives in Zr isotopes are placed between
the results obtained from the oblate and prolate coexistent shapes that
appear very close in energy in the PECs. The predicted half-lives in
the heavier Zr isotopes  $^{108,110}$Zr, where there are no experimental data
yet, are however very close to each other for both oblate and prolate shapes
and much larger than the predictions from the spherical shapes.
On the other hand, the measured half-lives in Mo isotopes
agree better with the calculations from oblate shapes, which are
lower than the corresponding prolate ones. Once more, in the heavier
isotopes $^{112,114}$Mo, the predicted half-lives for both shapes are
very close and larger than the spherical ones.
Thus, comparison with experimental half-lives indicates that in some
cases (Mo isotopes) a single shape accounts for this information, while
in other cases (most of the Zr isotopes) a more demanding treatment in
terms of mixing of different shapes seems to be more appropriate.
$P_n$ values are in general not well reproduced, although experimentally
only upper limits are measured in most cases.
Hence, it will be certainly worth measuring those heavier isotopes
and check whether they are properly described by the deformed shapes, or
a spherical component is needed as well.

Nevertheless, one should keep in mind that half-lives are integral properties
that collect all the information of the decay in a single number and does not
tell us about the detailed internal structure of the GT strength distribution,
much more sensitive to the nuclear structure.
From a more detailed analysis of the GT strength distributions in the
$Q_\beta$ energy range accessible in $\beta$-decay, we have shown that
the differences between the predictions of the different nuclear shapes
could be clearly distinguished experimentally.
Although these spectroscopic measurements are at present not feasible
because of the still low production rates of exotic nuclei at modern
radioactive beam facilities, they will provide in the future precise tests
of the nuclear structure in exotic nuclei.

\begin{acknowledgments}
This work was supported by Ministerio de Ciencia e Innovaci\'on
(Spain) under Contract No. FIS2008--01301. It was also
supported in part by the Joint Institute for Nuclear 
Astrophysics (JINA) under NSF Grant PHY-02-16783 and the National 
Superconducting Cyclotron Laboratory (NSCL) under NSF Grant PHY-01-10253.

\end{acknowledgments}

\end{document}